\documentstyle[preprint,aps,epsfig]{revtex}

\title{A phenomenological spin--orbit three-body force}
\author{A. Kievsky\thanks{ Permanent address: Istituto Nazionale di 
        Fisica Nucleare, Via Buonarroti 2, 56100 Pisa, Italy}}
\address{Department of Physics and Astronomy, University of North Carolina
 at Chapel Hill, NC 27599-3255, USA and Triangle Universities Nuclear
 Laboratory, Durham, NC 27708, USA}

\begin{document}

\maketitle
\begin{abstract}
 In order to improve the description of the N-d vector analyzing powers 
 $A_y$ and $iT_{11}$ (the so-called $A_y$ puzzle) a spin--orbit three-body
 force has been introduced. The ${\bf L}\cdot{\bf S}$ term in the 
 $NN$ potential has been modified including a two-parameter three-body 
 function depending on the hyperradius $\rho$. The modification has
 been performed in channels where the pair of nucleons $(i,j)$ are coupled 
 to spin $S_{ij}=1$ and isospin $T_{ij}=1$. Calculations have been
 done in the energy region from $E_{lab}=648$ keV up to $E_{lab}=10$ MeV.
 A noticeable improvement in the description of the polarization
 observables has been obtained.

\end{abstract}

\section{Introduction}

In the standard description of light nuclei as composed of structureless
nucleons, three--nucleon forces (3NF's) play an important role. The simple
assumption of nucleons interacting through a pairwise $NN$ potential
fails to reproduce the experimental binding energies. This  suggests that
the model has to be extended to include many--body forces in the potential 
energy. 
First derivations of 3NF's are based on two-pion exchange involving
a $\Delta$ excitation.
The most often used potentials of this kind are the Tucson-Melbourne 
(TM)~\cite{TM}, the Brazil (BR)~\cite{BR} models,
and the Urbana (UR)~\cite{UR}
model which also includes a central phenomenological repulsive term
with no spin-isospin structure. In practical applications, the chosen
3NF is adjusted to reproduce the $A=3,4$ binding energies. 
When the calculations are extended to describe bound states in 
$p$--shell nuclei, a persistent underbinding is observed in the mass
region with $A=5-8$~\cite{A8}.
In particular, several excited states are not well reproduced indicating
that the 3NF could contain a more complicate structure.

A different problem has been observed in N-d scattering at low energies.
The calculated vector analyzing powers 
$A_y$ and $iT_{11}$ show an unusual
large discrepancy with the experimental data~\cite{ay1,KRTV96}.
Attempts to improve the description of these observables with the
inclusion of a two-pion three-nucleon force ($2\pi$--3NF)
were unsuccessful~\cite{ay1,KRTV96,KVR95,report}.  
This discrepancy has been called the $A_y$ puzzle since it is, 
by far, the largest disagreement observed in the theoretical 
description of the three-nucleon system at low energies.
Recently, it has been shown that the puzzle is not limited to
the three-nucleon system but a similar problem appears in the
calculation of $A_y$ in $p$-$^3$He scattering~\cite{viv1}. 
The $A_y$ puzzle could be also 
a signal for different forms in the three-body potential~\cite{carl,hub98}

In ref.~\cite{KRTV96}
p-d and n-d scattering have been studied below the deuteron breakup.
Using the experimental data for cross section
and vector and tensor analyzing powers at $E_{lab}=2.5$ and 
$3.0$ MeV from Shimizu et al.~\cite{shimizu}, 
it was possible to perform a phase-shift analysis
(PSA) and to compare the results to the theoretical phases. The conclusion
was that small differences in the $P$-wave parameters are responsible
for the large disagreement in $A_y$ and $iT_{11}$. 
It is a general feature that all realistic $NN$ potentials underpredict the
splitting in the $^4P_J$ phases and the magnitude of the
$\epsilon_{3/2}-$ mixing parameter.
When one of the $2\pi$--3NF's is included in the Hamiltonian
there is not an appreciable reduction of the discrepancy, 
giving in some cases a poorer description. In fact, the inclusion of
the $2\pi$--3NF tends to reduce the splitting in $^4P_J$ and to
slightly increase $\epsilon_{3/2}-$. These two opposite effects almost cancel 
each other in the construction of $A_y$ and $iT_{11}$.
The operator's form of these particular models of 3NF's does not include 
${\bf L}\cdot{\bf S}$ terms
which are to a large extent responsible for the splitting in the
$P$-waves parameters.

In the present paper a phenomenological spin--orbit three--nucleon force 
(SO--3NF) is introduced in order to study its effect on the N-d
vector analyzing powers $A_y$ and $iT_{11}$ at low energies. 
The ${\bf L}\cdot{\bf S}$ term in the $NN$ potential has been
modified including a two-parameter three-body 
function depending on the hyperradius $\rho$. The two parameters are related 
to the strength and range of the force and they have been fixed with
the intention to improve the
description of $A_y$ and $iT_{11}$ at $E_{lab}=3.0$ MeV, just below 
the deuteron breakup. Three different sets of parameters have been
considered and, accordingly, used to calculate
scattering observables from $E_{lab}=648$ keV up to $E_{lab}=10$ MeV.
The SO--3NF has been introduced in channels where the pair of
nucleons $(i,j)$ are coupled to spin $S_{ij}=1$ and isospin $T_{ij}=1$.
At the level of the two-nucleon ($2N$) system this channel is related to
scattering in odd waves. In N-d scattering the $^4P_J$ parameters and hence
$A_y$ and $iT_{11}$ are very sensitive to the force in this 
particular channel.
On the other hand its effect on
the binding energy of the three-nucleon system is very small.
This is the opposite behavior from that produced by the $2\pi$--3NF 
which gives the main contribution in the $J=1/2^+$ state and has
small influence on the vector observables. Therefore these
two different classes of 3NF's are to some extent complementary 
and can be studied separately. 

The calculations of the N-d scattering observables presented here
in the following at different energies have been done using
the Pair Correlated Hyperspherical Harmonic (PHH)~\cite{KVR93} basis.
In this method the wave function of the system is expanded in
terms of correlated basis elements and
the description of the system proceeds via a variational principle.
Bound states are obtained using the Rayleigh-Ritz variational principle
whereas scattering states are obtained using the generalized Kohn
variational principle. This technique has been extensively discussed in
refs.~\cite{KVR95,KVR94,kie97} for energies below the deuteron breakup 
threshold and, very recently, also applied to energies above the
breakup threshold~\cite{KVR97,KVR98}.

The paper is organized as follows. In Sec. II the 
two parameter spin--orbit three-body force is introduced. In Sec. III
the polarization observables as well as the binding energy of $^3$He
are studied for specific values of these parameters.
The conclusions and perspectives are given in the last section.

\section{A two-parameter spin--orbit three-body force}

Disregarding for the moment the presence of three-nucleon forces,
the potential energy operator of the three-nucleon system is
\begin{equation}
   V_{3N}=\sum_{i<j}V_{2N}(i,j) \ ,
\label{eq:V3N}
\end{equation}
where
$V_{2N}$ is the $NN$ interaction that, in general, is constructed by
fitting the $2N$ scattering data and the deuteron properties. Recently,
several potential models have been determined including explicitly
charge dependence which describe the $2N$ data with a
$\chi^2$ per datum $\approx 1$. 
Here we will refer to the Argonne AV18 interaction, which is one
of these new generation potentials~\cite{AV18}. The nuclear part of 
the AV18 potential consists of a sum over 18 different terms. The first
14 terms are charge independent whereas the four additional operators
introduce charge symmetry breaking.  Each of the first 14 terms 
includes a projector $P_{ST}(ij)$ onto the spin-isospin 
states $S,T$ 
of particles $(i,j)$ multiplied by one of the following operators,
${\cal O}^p=1,S_{12},{\bf L}\cdot{\bf S},L^2,({\bf L}\cdot{\bf S})^2$.
The strength of each term is given
by a scalar function $v^p_{ST}(r_{ij})$ depending
on the relative distance between particles $(i,j)$.
For example the ${\bf L}\cdot{\bf S}$ interaction between particles $(i,j)$
is defined in the two 
channels with isospin $T_{ij}=0,1$ and spin $S_{ij}=1$ and is given by
the functions $v^{ls}_{10}(r_{ij})$ and $v^{ls}_{11}(r_{ij})$,
respectively. 

In the three-nucleon system odd-parity scattering states are directly 
related to the potential in channels where particles $(i,j)$ are coupled to
$S_{ij}=1$, $T_{ij}=1$. In particular here we are interested in the
${\bf L}\cdot{\bf S}$ term and its relation to the vector observables
$A_y$ and $iT_{11}$.
Limiting the discussion to the ${\bf L}\cdot{\bf S}$ interaction in this
particular channel, the corresponding term in
the three-nucleon potential energy is
\begin{equation}
   V^{ls}_{3N}=
   \sum_{i<j}v^{ls}_{11}(r_{ij}){\bf L}_{ij}\cdot{\bf S}_{ij}P_{11}(ij) \ .
\end{equation}

 At this point
 we can conjecture that the presence of the third nucleon (particle $k$)
 could modify the interaction, introducing in the function $v^{ls}_{11}$ 
 a dependence on the distances between the third nucleon $k$ and 
 particles $(i,j)$.
 Accordingly, the above interaction transforms to
\begin{equation}
   V^{ls}_{3N}=
   \sum_{i<j}{1\over 2}[w^{ls}_{11}(r_{ijk}){\bf L}_{ij}\cdot{\bf S}_{ij}
            +{\bf L}_{ij}\cdot{\bf S}_{ij}w^{ls}_{11}(r_{ijk})]P_{11}(ij) \ ,
\end{equation}
 where $r_{ijk}$ is a scalar function of the three interparticle distances
 $r_{ij},r_{jk},r_{ki}$.
 The symmetric form has been introduced since, in general, the
 ${\bf L}\cdot{\bf S}$ operator does not commute with an operator depending
 on $r_{ijk}$. Different forms are possible for
the three-body interaction $w^{ls}_{11}(r_{ijk})$ provided that
$w^{ls}_{11}(r_{ijk})\rightarrow v^{ls}_{11}(r_{ij})$ when 
$r_{ik},r_{jk}\rightarrow\infty$.
A simple two-parameter form that we are going to analyze here is
\begin{equation}
   w^{ls}_{11}(r_{ijk})=v^{ls}_{11}(r_{ij})+W_0{\rm e}^{-\alpha\rho} \ ,
\label{eq:twop}
\end{equation}
where the hyperradius $\rho$ is
\begin{equation}
   \rho^2={2\over 3}(r_{12}^2+r_{23}^2+r_{31}^2)
\end{equation}
and $W_0$ and $\alpha$ are parameters characterizing the strength and range
of the three-body term. When the dependence in the
scalar function $r_{ijk}$ is limited to $r_{ij}$ and $\rho$, the operators
$w^{ls}_{11}(r_{ij},\rho)$ 
and ${\bf L}_{ij}\cdot{\bf S}_{ij}$ commute. Accordingly,
the spin--orbit force becomes
\begin{equation}
   V^{ls}_{3N}= \sum_{i<j}v^{ls}_{11}(r_{ij})
          {\bf L}_{ij}\cdot{\bf S}_{ij}P_{11}(ij)+
          W_0{\rm e}^{-\alpha\rho}\sum_{i<j}
          {\bf L}_{ij}\cdot{\bf S}_{ij}P_{11}(ij) \ .
\label{eq:ls3}
\end{equation}
Replacing this term in eq.(\ref{eq:V3N}) and including now also the 
$2\pi$-3NF, the final form for the three-nucleon potential energy 
to be used in the present work is
\begin{equation}
   V_{3N}=\sum_{i<j}V_{2N}(i,j)+\sum_{i<j<k}W^{ls}_{3N}(i,j,k) 
         +\sum_{i<j<k}W^{2\pi}_{3N}(i,j,k) \ .
\label{eq:V3NF}
\end{equation}
The $W^{ls}_{3N}$ term is the phenomenological spin--orbit force 
defined in the second term of eq.(\ref{eq:ls3}); for the $W^{2\pi}_{3N}$
term the discussion will be limited to the Urbana force.

The choice of the two-parameter exponential form in the
definition of $w^{ls}_{11}$ (see eq.(\ref{eq:twop})) is arbitrary and was
selected in order to make a phenomenological representation of a 3NF
containing a spin--orbit interaction.
The hyperradial dependence is the simplest scalar
function depending on the three-interparticle distances which has
the property of commuting with the spin--orbit operator. These choices,
driven by simplicity, have been made in order to focus on 
the spin--orbit operator and its relation to $A_y$ and $iT_{11}$
in p-d scattering. The numerical values for the constants $W_0,\alpha$
are discussed in the next section.

In principle, the argument used to introduce the scalar 
function $w^{ls}_{11}(r_{ijk})$ as a modification of the function
$v^{ls}_{11}(r_{ij})$ due to the presence of particle $k$, could be 
extended to the other functions $v^p_{ST}(r_{ij})$ of the $NN$ potential. 
This leads to a three-body force with an hybrid form
in which nucleons $(i,j)$ interact with the
same operator's structure of the $NN$ potential but with scalar functions
depending on the interparticle distances of the three nucleons $(i,j,k)$. 
The original function $v^p_{ST}(r_{ij})$ must be recovered when nucleon $k$
is at $\infty$.  In the present work we
are limiting the argument to one specific term and
trying to relate its parametrization
to the two vector observables $A_y$ and $iT_{11}$.

\section{N-d scattering calculations with the SO--3NF}

The inclusion of the SO--3NF has been done with the hope of
improving the description of $A_y$ and $iT_{11}$
without destroying the agreement already observed for
the cross section and tensor observables. At energies below the deuteron
breakup, measurements for tensor and vector observables exist
for p-d scattering at $E_{lab}=648$ keV~\cite{carl} and at
$E_{lab}=2.5$ MeV and $3.0$ MeV~\cite{shimizu,knutson}. 
The $648$ keV data set is at
the lowest energy at which these measurements have been made.
The other two sets of measurements lie just below the deuteron breakup. 
Theoretical calculations
using the AV18 potential underpredict $A_y$ and $iT_{11}$ of about $30\%$.
Comparisons to phase-shift and mixing parameters extracted from
PSA performed at these three energies~\cite{KRTV96,carl,knutson}
show the aforementioned insufficient splitting in the $^4P_J$ phase shifts
as well as an underpredicted  $\epsilon_{3/2}-$.
Calculations using the AV18+UR potential
essentially do not change these findings. The almost constant
underprediction in the vector polarization observables (in percentage)
and the fact that the inclusion of the $2\pi$--3NF's does not increase the
splitting in $^4P_J$ is a motivation for considering
new additional forms for the three--body potential.
The selection of the ${\bf L}\cdot{\bf S}$ operator
is a natural choice since when applied to the $^4P_J$ state it acts with
opposite sign in the states $J=1/2^-$ and $J=5/2^-$, tending to increase
the splitting.

Three different choices of the exponent $\alpha$ in the hyperradial
spin--orbit interaction defined in eq.(\ref{eq:twop})
have been selected with the intention of constructing forces
with different ranges.  The strength $W_0$ has been adjusted in each case 
in an attempt to improve the description of the vector observables. 
The analysis has been performed at $E_{lab}=3.0$ MeV.
The selected ranges are $\alpha=0.7,1.2,1.5$ fm$^{-1}$, so as to
simulate a long, medium and short range force.
The corresponding values for the depth are $W_0=-1,-10,-20$ MeV. 
The calculations have been performed using the nuclear part of
AV18 plus the Coulomb interaction.
The $2\pi$-3NF has been disregarded at the present
stage since its contribution to the description of the vector observables
is small. 

The results for the proton and neutron analyzing powers $A_y$
and the deuteron analyzing power $iT_{11}$ are given in Fig.1 together
with the experimental data of ref.~\cite{shimizu,nay}. The four
curves correspond to the AV18 potential and the three
different choices for the parameters $(\alpha,W_0)$. 
The dotted line is the AV18 prediction and shows the expected discrepancy. 
The solid line corresponds to the AV18 plus the long range force 
(AV18+LS1), the long-dashed line to the AV18 plus the medium range force
(AV18+LS2) and the dotted-dashed line to the AV18 plus the short range force
(AV18+LS3). The inclusion of the spin--orbit force improves the
description of the vector observables,
although there is a slightly different sensitivity
in $A_y$ and $iT_{11}$. The AV18+LS1 curve is slightly above the data,
especially for $iT_{11}$. The AV18+LS2 curve is slightly below (above) the data
in $A_y$ ($iT_{11}$). The AV18+LS3 curve is
slightly below the data especially for $A_y$. 
In the bottom panel of Fig.1 the n-d analyzing power has been 
calculated using the same potential models as before. 
Again, there is an improvement in the description of $A_y$
equivalent to that one obtained in the p-d case. 
In Fig.2 the tensor analyzing powers $T_{20}, T_{21}, T_{22}$
are shown at the same energy and compared to the data
of ref.~\cite{shimizu}. The inclusion of the SO--3NF has no
appreciable effect and the four curves are practically on top of each other.
These observables are not very sensitive to the splitting in $^4P_J$-waves.
They are sensitive to scattering in $D$--waves and higher partial waves,
which are only weakly distorted by the SO--3NF.

Before extending the calculations to other energies the analysis of the
binding energy of $^3$He deserves some attention. We expect that the
inclusion of the $W^{ls}_{3N}$ term will produce only
a small distortion in the
bound state due to the low occupation probability of channels with
$S_{ij}=1$ and $T_{ij}=1$. 
This is corroborated by the calculations shown in table I.
The binding energy, the kinetic energy and the $S'$-, $D$-, and $P$-wave
probabilities are given for the different potential models. 
The AV18 and the AV18+UR interactions have been considered. The 
$2\pi$--3NF of Urbana has been now taken into account since it produces large
effects in the bound state. Calculations have been done for 
these two potentials with and without the inclusion
of the three different choices for the SO--3NF. Its inclusion
produces a small repulsion which in any case is not
greater than $50$ keV, with all the other mean values modified very little.  
The force with the longest range produces very tiny
effects due to its very small strength. The other two forces produce
slightly greater, and similar, modifications.
We can conclude that the structure of the three-body
bound state remains essentially unaffected by the spin--orbit 3NF with
the ranges and strengths considered.
 
The analysis of the bound state is important since changes
in $A_y$ and $iT_{11}$ of the size shown in Fig.1 could, in
principle, be obtained with modified forms of the $2\pi$--3NF's. 
But, in general, these modifications are not anymore compatible with 
a correct description of the bound state.
This is not the case
for the spin--orbit 3NF we are considering. Therefore, we can extend the
calculations to other energies based in the fact that the new model
makes a selective and appreciable effect only in the vector
observables. Calculations have been done at
$E_{lab}=648$ keV and $2.5$ MeV. The results for $A_y$ and $iT_{11}$
are shown in Fig.3 and compared to the experimental data of
refs.~\cite{carl,shimizu}.
Again a remarkable improvement in the description of both observables
is obtained. At the lowest energy the AV18+LS1 model seems to be
more effective because of its long range whereas for the other two
ranges the centrifugal barrier plays some role. The results
at $E_{lab}=2.5$ MeV have the same characteristics as those at $3.0$ MeV.

In table 2 we compare results for the $P$-waves parameters at 
$E_{lab}=2.5$ and $3.0$ MeV. 
Again, different cases using the AV18 potential with 
and without the inclusion of the spin--orbit 3NF have been considered.
In the last column the parameters corresponding
to the PSA from ref.~\cite{KRTV96} are given for the sake of comparison.
In the three cases the SO--3NF increases
the splitting of the $^4P_J$ phases. The phases $^4P_{1/2}$ and
$^4P_{5/2}$ and the mixing parameter 
$\epsilon_{3/2-}$, which play a major role
in the description of the vector observables, are now in much better
agreement with the PSA values. It was not obvious from the beginning
that with two parameters $(\alpha,W_0)$, it would be possible 
to increase the difference $\Delta P= ^4P_{5/2}-^4P_{1/2}$
in the measure suggested by the PSA, together with a
change in $\epsilon_{3/2-}$ in the expected direction and magnitude.
At the three energies the change 
in $\epsilon_{3/2-}$ was observed to happen in the correct direction.
Not all parameters are closer to the PSA values after including the 
SO--3NF.
For example $^4P_{3/2}$ has slightly changed in the opposite direction to 
that suggested by the PSA. But the final result in the description of the 
observables was found always to produce an improvement.

In order to complete the study of the SO-3NF
the extension to p-d calculations above the breakup channel is now
considered.
This extension is not straightforward since when the breakup channel is open
a correct description of the three outgoing nucleons has to be done.
In particular, in the p-d case, the Coulomb interaction introduces
difficulties that have been, and are at present, 
subject of intense investigation.
Recently the PHH technique has been extended to describe p-d elastic
scattering above the deuteron breakup~\cite{KVR97,KVR98}. The method is based
on the use of the Kohn variational principle in its complex form~\cite{kie97}
and it provides an accurate description of the polarization observables.
In ref.~\cite{KVR98} differential cross sections and vector and tensor 
analyzing powers at $E_{lab}=5$ and $10$ MeV have been calculated using the 
AV18 interaction. In the present work, calculations have been done at the
same two energies using the
AV18 interaction with and without the SO--3NF. The results for
$A_y$ and $iT_{11}$ are given in Fig.4 together with the experimental
data from ref.~\cite{sagara} ($E_{lab}=5$ MeV) and ref.~\cite{sperisen}
($E_{lab}=10$ MeV). Also at these energies we observe the same trend
as before, both observables are now better described. The sensitivity
to the different ranges is slightly different with the medium
range model starting to be more effective. In fact, at $5$ MeV 
the long and medium range curves overlap and at $10$ MeV
the splitting in the curves is now reversed in that the upper
curve corresponds to the medium range force. 

Finally, let us extend the analysis of the effects of the SO--3NF to other
observables at different energies. In Fig.5 the effects on the elastic 
cross section are shown in the energy interval from $E_{lab}=648$ keV up to
$E_{lab}=10$ MeV. At each energy, the four curves corresponding to the 
different potential models are on top of each other. 
There is not enough sensitivity in this
observable to changes in the phase-shift and mixing parameters of
the magnitude introduced by the SO--3NF used. 
The tensor analyzing powers at $E_{lab}=5$ MeV and $10$ MeV are shown
in Fig.6. The effects on these observables are very small. The three
curves corresponding to the AV18 plus the SO--3NF are practically
superposed with slight differencies with the AV18 curve (dotted line).
In particular, the second
minimum of $T_{20}$ is slightly improved as well as the second
maximum of $T_{21}$, whereas $T_{22}$ is essentially unchanged.

In order to give a quantitative measure of the improvement introduced by
the SO-3NF in the description of the data, a $\chi^2$ (per datum)
analysis is displayed in table III. The analysis includes the cross
section, the two vector and the three tensor analyzing powers at
$E_{lab}=648$ keV, $3$ MeV, $5$ MeV and $10$ MeV. The data set has been
taken from ref.~\cite{carl,shimizu,sagara,sperisen,carl1}. The $\chi^2$
per datum has been calculated using the theoretical predictions of
the AV18 potential model with and without the inclusion of the SO-3NF.
The cases corresponding to the three different choices of the strength
and range parameters have been considered.
From the table is clear the selective effect of the spin-orbit force on
the vector observables. There is a dramatic improvement in terms of
$\chi^2$ in these observables, whos value is reduced by more than one 
order of magnitude in several cases. At each energy,
the AV18+LS1 and AV18+LS2 potential models give the best description.
At the first three energies both potential models reproduce the data 
reasonably well and with similar quality.
At $10$ MeV the obtained $\chi^2$ per datum for $A_y$ is very high 
in all cases. This is a consequence of the extremely small error bars
in the data set at this particular energy. However, the $\chi^2$ 
has been improved by a factor of 
$\approx$ 25 going from AV18 to AV18+LS2 and a further reduction could 
be obtained with a fine tune of the strength and range parameters 
$W_0,\alpha$.

Looking at the differential cross section, in all cases the $\chi^2$ 
per datum is a large number and it changes very little when the SO-3NF 
is included. There is a sensitivity to the $^3$He binding energy which 
is not well reproduced unless a $2\pi$-3NF is considered.  For example, 
the value $\chi^2=30.3$ at $3$ MeV obtained with the AV18 potential
reduces to $\chi^2=4.0$ with the AV18+UR potential.
In the case of the tensor observables the changes in terms of $\chi^2$
when the SO-3NF is considered are moderated. There is a slight improvement 
in the description of $T_{20}$ and $T_{21}$, whereas the reverse situation 
is seen in $T_{22}$.

\section{Conclusions}

Elastic N-d scattering has been studied in the energy range from
$E_{lab}=648$ keV to $E_{lab}=10$ MeV using a potential model which
includes a spin--orbit three-nucleon interaction. This three-body potential
was introduced as a ``distortion'' of the function $v^{ls}_{11}(r_{ij})$
of the $NN$ potential. This function gives the magnitude of the 
${\bf L}\cdot{\bf S}$ interaction
in the channels where particles $(i,j)$ are coupled to spin $S_{ij}=1$ and
isospin $T_{ij}=1$ and was converted to a function $w^{ls}_{11}(r_{ijk})$
depending on the three-interparticle distances $r_{ij},r_{jk},r_{ki}$.
The condition on $w^{ls}_{11}$ was that the interaction $v^{ls}_{11}$
is recovered when the third particle is far from the other two.
A phenomenological
hyperradial exponential form depending on two parameters was used which
fixes the range and strength of the three-body part of the interaction. 

The choice of the ${\bf L}\cdot{\bf S}$ operator acting on the 
$S_{ij}=1$, $T_{ij}=1$ spin-isospin channels was based on the large 
effects it has in the description of the two vector observables
$A_y$ and $iT_{11}$ in N-d scattering. The origin of this discrepancy 
lies in the too low splitting predicted by all realistic potential
models for the $^4P_J$ phase shifts. There is a close relation, 
due to the Pauli blocking,
between scattering in $^4P_J$--waves and the interaction in this channel.
Moreover the ${\bf L}\cdot{\bf S}$ operator is attractive (repulsive)
in the $J=1/2^-$ state ($J=5/2^-$ state), therefore increasing the splitting
of these phases.

The study of the SO--3NF was done phenomenologically by
fixing three different values for its range parameter
$\alpha$ and selecting the strength parameter $W_0$ to provide
a better description of the vector observables. The energy at
$E_{lab}=3.0$ MeV was chosen for this analysis and successively the
three sets of parameters were used to describe the same observables
at different energies. The difference between the curves obtained for the
description of $A_y$ and $iT_{11}$ after including the spin--orbit force
could be further reduced with a fine tuning of $W_0$.
An important check was made resulting in the observation that the
structure of the bound state and other observables, for example the
tensor analyzing powers are not appreciable disturbed by this
interaction. Moreover the n-d and p-d $A_y$ are described equally well 
when the spin--orbit 3NF is included. This means that no additional 
charge--symmetry breaking effects are needed.

The use of this phenomenological SO--3NF improves the description
of the vector observables in all the studied cases, from
$E_{lab}=648$ keV to $E_{lab}=10$ MeV. The sensitivity
of the observables to the range parameters is slightly different at the
energies studied and there is not a definitive preference for one of
them. Perhaps the set with the medium range parameter 
($\alpha=1.2$ fm$^{-1}$, $W_0=-10$ MeV) gives on average the best
description. A better evaluation about the simple hyperradial
dependence introduced here because of its nice property of commuting
with the ${\bf L}\cdot{\bf S}$ operator, can be obtained only after
extending the calculations to higher energies. The main conclusion
of the present work is the identification of the ${\bf L}\cdot{\bf S}$ 
force in the $S_{ij}=1$, $T_{ij}=1$ spin-isospin channel as one which
can resolve the $A_y$ puzzle. In addition, a simple 
model has been proposed to repair the discrepancy. 

Further investigations, which are in progress, consist in the inclusion
of the $2\pi$--3NF in the description of the scattering observables
other than in the bound state, the extension of the calculations
to higher energies and the study of the force in the four--body
reaction p-$^3$He. Whereas in the first case we can expect at most
a small change of the two parameters $(\alpha,W_0)$, the two other
studies will give more insight about the force. In particular, by
studying the p-$^3$He reaction the extension to heavier systems
can be tested. 

Finally, the possibility of introducing a dependence on the three
interparticle distances in other
NN functions $v^p_{ST}(r_{ij})$, as discussed at the end of Sec. II,
deserves some attention. Disregarding other types of
3NF's, the modified potential energy between three nucleons would be
\begin{equation}
   V_{3N}= \sum_{i<j}
           \sum_{S,T}\sum_p{1\over 2}[w^p_{ST}(r_{ijk}){\cal O}^p_{ij}
            +{\cal O}^p_{ij}w^p_{ST}(r_{ijk})]P_{ST}(ij) \ .
\end{equation}
The original functions $v^p_{ST}$ are recovered
when nucleon $k$ is far from the pair $(i,j)$. In addition, the symmetric
form can be avoided if the three-body radial dependence is limited to the
hyperradius, i.e. $w^p_{ST}(r_{ijk})=w^p_{ST}(r_{ij},\rho)$.
The next step is the parametrization of the
functions $w^p_{ST}$ with a number of
parameters to be determined by a fit procedure. The maximum
number of free parameters and the selection of the channels where
the distorted functions are introduced are related to the observables
included in the fit. For example, it would be impossible to reproduce
simultaneously the binding energy of $^3$He and the vector analyzing
powers in p-d scattering by modifying the potential 
only in channels with $S_{ij}=1,T_{ij}=1$
and neglecting the $2\pi$-3NF. Conversely, this would be possible
by extending the modification to other terms in channels with 
$S_{ij}=1,T_{ij}=0$ or $S_{ij}=0,T_{ij}=1$. 
 
\begin{acknowledgements}
I would like to thank the University of North Carolina at
Chapel Hill and the Triangle Universities Nuclear Laboratory for
hospitality and support during my stay in Chapel Hill, where this work
was performed. Moreover, I would like to thank W. Tornow, E. Ludwig, 
H. Karwowski, C. Brune, L. Knutson and E. George 
as well as M. Viviani, S. Rosati and A. Fabrocini for useful discussions.
\end{acknowledgements}

\newpage 

Table Captions

Table 1. The binding energy of $^3$He is shown together with the
kinetic energy and the $S'$-, $P$- and $D$-wave probabilities. The
AV18 and the AV18+UR potentials are considered with and without the
spin--orbit 3NF for three choices for the parameters
(see text).

Table 2. The $P$-waves phase shift and mixing parameters calculated
at two different energies. The AV18 potential and 
three different sets for the parameters in the
spin--orbit 3NF (see text) have been used. The results from the PSA of 
ref.~\cite{KRTV96}
are given in column 6 for the sake of comparison.

Table 3. $\chi^2$ per datum at four different energies calculated
using the AV18 potential model with and without the three different
parametrization of the SO-3NF. The data set has been taken from
refs.~\cite{carl,shimizu,sagara,sperisen,carl1}. The number in parenthesis
is the number of data points.

\newpage 

Figure Captions

Fig.1. The p-d and n-d analyzing power $A_y$ and the deuteron analyzing
power $iT_{11}$ are shown at $E_{lab}=3.0$ MeV. The different curves
correspond to the following potential models: AV18(dotted line),
AV18+LS1 (solid line), AV18+LS2 (dashed line), AV18+LS3 (dotted-dashed line).
The experimental data are from ref.~\cite{shimizu} (p-d) and
ref.~\cite{nay} (n-d).

Fig.2. The tensor analyzing powers $T_{20}, T_{21}, T_{22}$ at
$E_{lab}=3.0$ MeV for the same potential model as Fig.1.
The experimental data are from ref.~\cite{shimizu}.

Fig.3. $A_y$ and $iT_{11}$ at two different lab energies.
The four curves correspond to the same potentials as in the previous
figures.  Experimental data are from ref.~\cite{carl} at $648$ keV and 
from ref.~\cite{shimizu} at $2.5$ MeV.

Fig.4. As in Fig.3 at two different energies above the deuteron
breakup threshold.  Experimental data are from ref.~\cite{sagara} 
at $5$ MeV and from ref.~\cite{sperisen} at $10$ MeV.

Fig.5. The differential cross section at five different lab energies. At each
energy the four curves corresponding to the AV18 potential with and
without the SO-3NF overlap. Experimental points are from
refs.~\cite{shimizu,sagara,carl1}.

Fig.6. The tensor analyzing powers at $E_{lab}=5$ MeV and $10$ MeV. At each
energy there are four curves corresponding to AV18 (dotted line), 
AV18+LS1 (solid line), AV18+LS2 (dashed line) and AV18+LS3 (dotted-dashed
line).

\newpage 

\begin{table}[h]
\begin{tabular} {c||c|c|c|c|c||}
\hline
\hline
 Potential  & B(MeV) & T(MeV) & $P_{S'}$(\%) & $P_P$(\%) & $P_D$(\%)  \\
\hline
 AV18       & -6.942 & 45.72  & 1.542        & 0.065     & 8.481      \\
 AV18+LS1   & -6.929 & 45.66  & 1.546        & 0.064     & 8.456      \\
 AV18+LS2   & -6.905 & 45.52  & 1.555        & 0.063     & 8.423      \\
 AV18+LS3   & -6.905 & 45.52  & 1.556        & 0.063     & 8.427      \\
\hline
 AV18+UR    & -7.768 & 50.25  & 1.251        & 0.132     & 9.262      \\
 AV18+UR+LS1& -7.751 & 50.17  & 1.255        & 0.131     & 9.233      \\
 AV18+UR+LS2& -7.718 & 49.99  & 1.264        & 0.129     & 9.190      \\
 AV18+UR+LS3& -7.718 & 49.97  & 1.265        & 0.129     & 9.194      \\
\hline                                     
\hline                                     
\end{tabular}
\caption{ }
\end{table}

\begin{table}[h]
\begin{tabular} {c||c|c|c|c|c||}
\hline
        $E_{lab}=3.0$ MeV               \\
\hline
 &AV18&AV18+LS1&AV18+LS2&AV18+LS3&PSA\\
\hline
 $^2P_{1/2}$& -7.36   & -7.40  & -7.38 & -7.37 & -7.41   \\
 $^2P_{3/2}$& -7.12   & -7.11  & -7.12 & -7.12 & -7.18   \\
\hline                                     
 $^4P_{1/2}$& 22.11   & 21.67  & 21.72 & 21.87 & 21.77   \\
 $^4P_{3/2}$& 24.23   & 24.10  & 24.14 & 24.18 & 24.30   \\
 $^4P_{5/2}$& 24.00   & 24.27  & 24.25 & 24.17 & 24.26   \\
\hline                                     
$\epsilon_{1/2-}$& 5.72 & 5.62 & 5.62 & 5.66 & 5.70 \\
$\epsilon_{3/2-}$&-2.22 &-2.37 &-2.33 &-2.29 &-2.46 \\
\hline                                     
        $E_{lab}=2.5$ MeV               \\
\hline
 &AV18&AV18+LS1&AV18+LS2&AV18+LS3&PSA\\
\hline
 $^2P_{1/2}$& -6.85   & -6.88  & -6.87 & -6.86 & -7.11   \\
 $^2P_{3/2}$& -6.70   & -6.67  & -6.68 & -6.68 & -6.90   \\
\hline                                     
 $^4P_{1/2}$& 20.10   & 19.70  & 19.75 & 19.87 & 19.89   \\
 $^4P_{3/2}$& 22.45   & 22.27  & 22.30 & 22.34 & 22.49   \\
 $^4P_{5/2}$& 21.84   & 22.08  & 22.07 & 22.00 & 22.14   \\
\hline                                     
$\epsilon_{1/2-}$& 4.95 & 4.87 & 4.87 & 4.90 & 4.77 \\
$\epsilon_{3/2-}$&-1.86 &-2.01 &-1.98 &-1.94 &-2.06 \\
\hline                                     
\end{tabular}
\caption{}
\end{table}

\begin{table}[h]
\begin{tabular} {c||c|c|c|c||}
\hline
$E_{lab}=648$ keV&AV18&AV18+LS1&AV18+LS2&AV18+LS3\\
\hline
 $\sigma$ (19)& 38.5 & 39.5 & 41.0 & 41.5 \\
 $A_y$     (3)& 22.7 & 0.4  & 2.6  & 8    \\
 $iT_{11}$ (3)& 18.3 & 3.4  & 1.4  & 4.3  \\
 $T_{20}$ (24)&  2.7 & 2.6  & 2.6  & 2.7  \\
 $T_{21}$ (22)&  3.9 & 3.9  & 3.9  & 3.9  \\
 $T_{22}$ (20)&  2.1 & 2.1  & 2.1  & 2.1  \\
\hline                                     
$E_{lab}=3$ MeV&AV18&AV18+LS1&AV18+LS2&AV18+LS3\\
\hline
 $\sigma$ (37)& 30.3 & 31.2 & 32.9 & 32.8 \\
 $A_y$    (38)&236.7 & 3.3  & 5.8  & 46.6 \\
 $iT_{11}$(51)&149.1 &10.4  & 3.9  & 14.8 \\
 $T_{20}$ (51)&  4.4 & 2.7  & 3.0  & 3.2  \\
 $T_{21}$ (51)&  6.8 & 5.0  & 4.4  & 4.5  \\
 $T_{22}$ (51)& 12.7 &14.8  &14.8  &14.3  \\
\hline                                     
$E_{lab}=5$ MeV&AV18&AV18+LS1&AV18+LS2&AV18+LS3\\
\hline
 $\sigma$ (41)& 29.3 & 30.2 & 30.2 & 30.5 \\
 $A_y$    (43)&229.8 & 6.9  & 8.5  & 38.5 \\
 $iT_{11}$(61)&120.1 & 2.7  & 3.2  & 5.3  \\
 $T_{20}$ (61)& 12.2 & 8.8  & 8.1  & 9.1  \\
 $T_{21}$ (61)& 11.2 & 9.3  & 8.6  & 9.2  \\
 $T_{22}$ (61)&  7.9 & 9.6  & 9.7  & 9.2  \\
\hline                                     
$E_{lab}=10$ MeV&AV18&AV18+LS1&AV18+LS2&AV18+LS3\\
\hline
 $\sigma$ (42)& 12.3 & 11.4 & 10.4 & 11.1 \\
 $A_y$    (48)&840.9 &80.1  &34.9  &128.3 \\
 $iT_{11}$(27)& 37.3 & 4.2  & 3.1  & 4.9  \\
 $T_{20}$ (27)& 12.8 & 8.4  & 6.7  & 7.9  \\
 $T_{21}$ (22)&  6.3 & 3.9  & 3.5  & 4.0  \\
 $T_{22}$ (27)& 25.4 &29.8  &32.0  &29.4  \\
\hline                                     
\end{tabular}
\caption{}
\end{table}

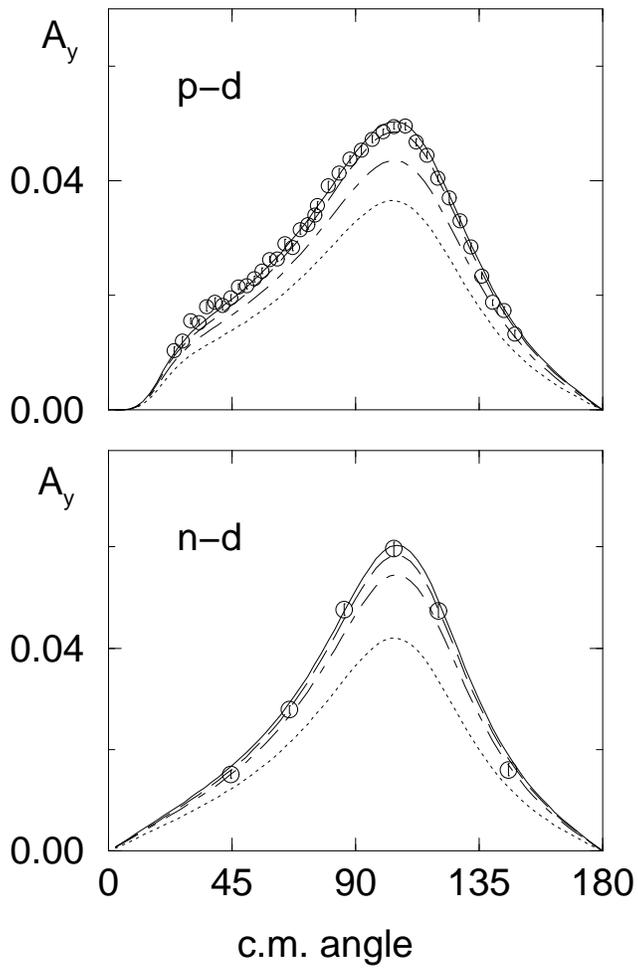
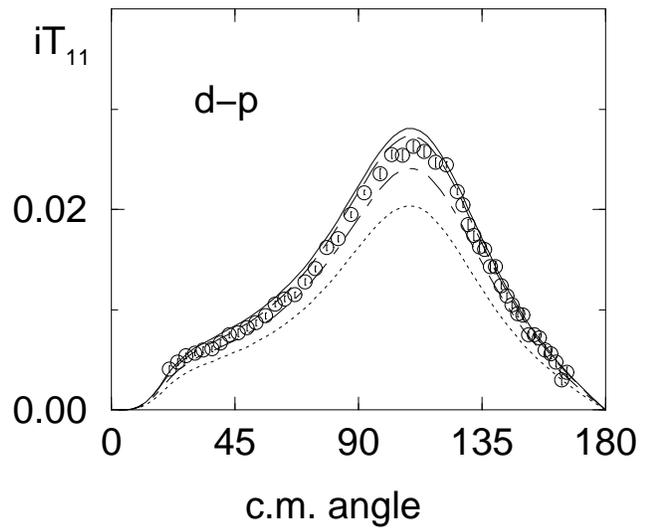
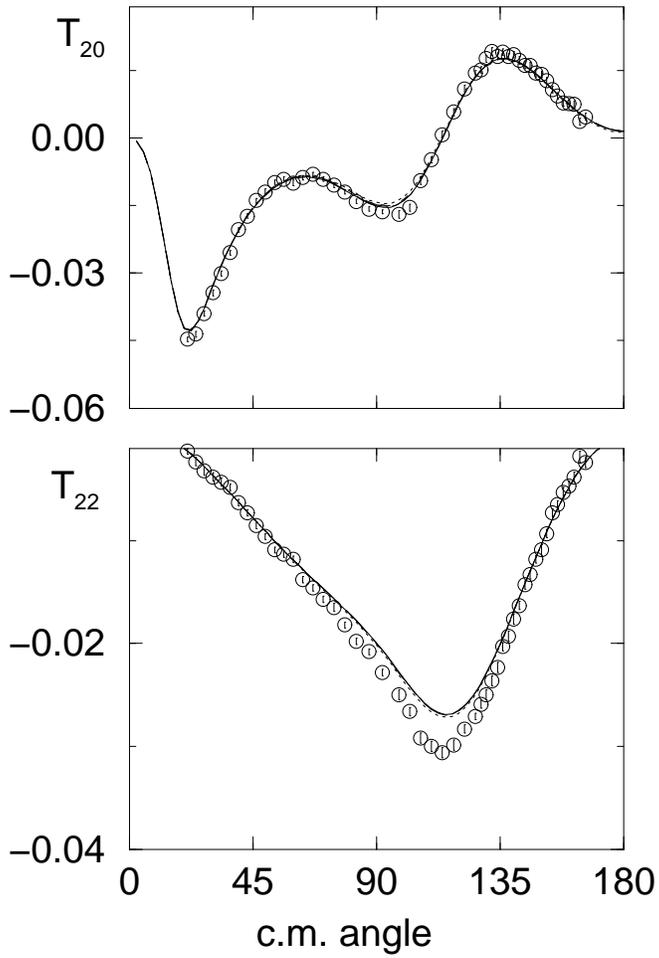
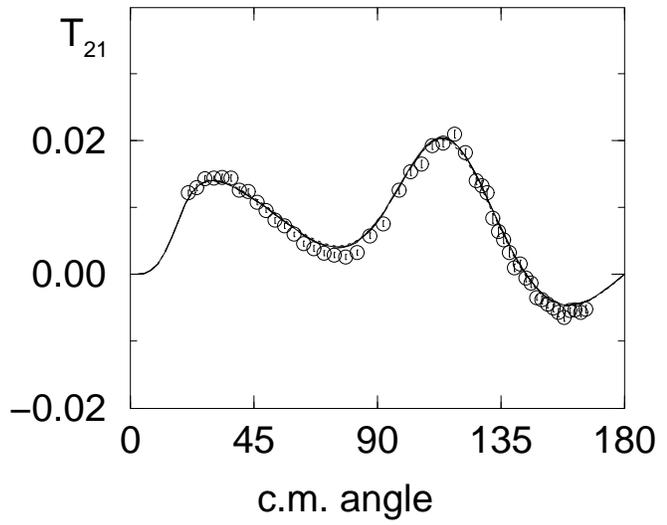
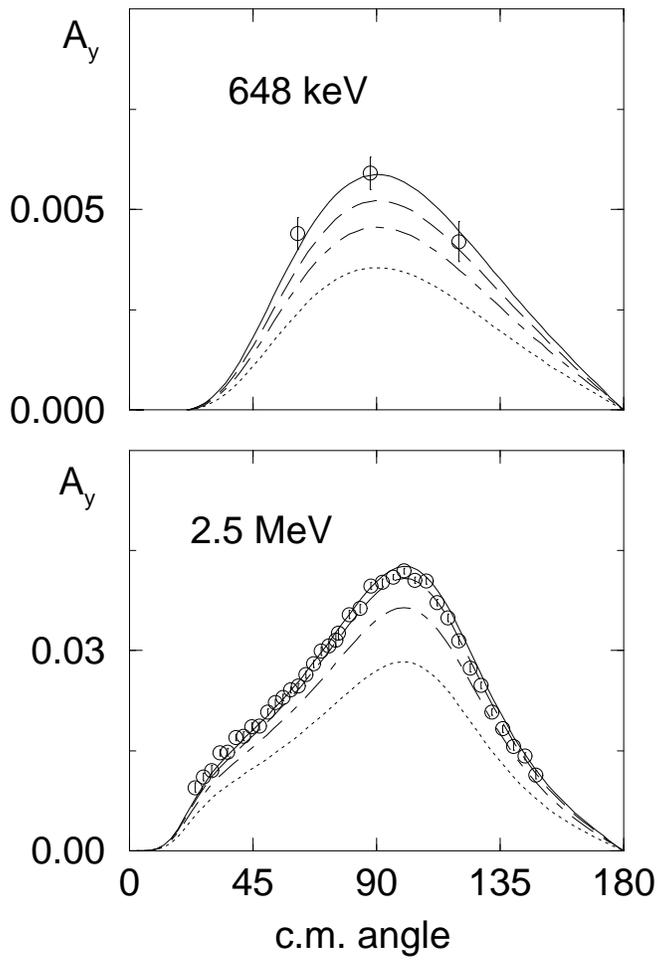
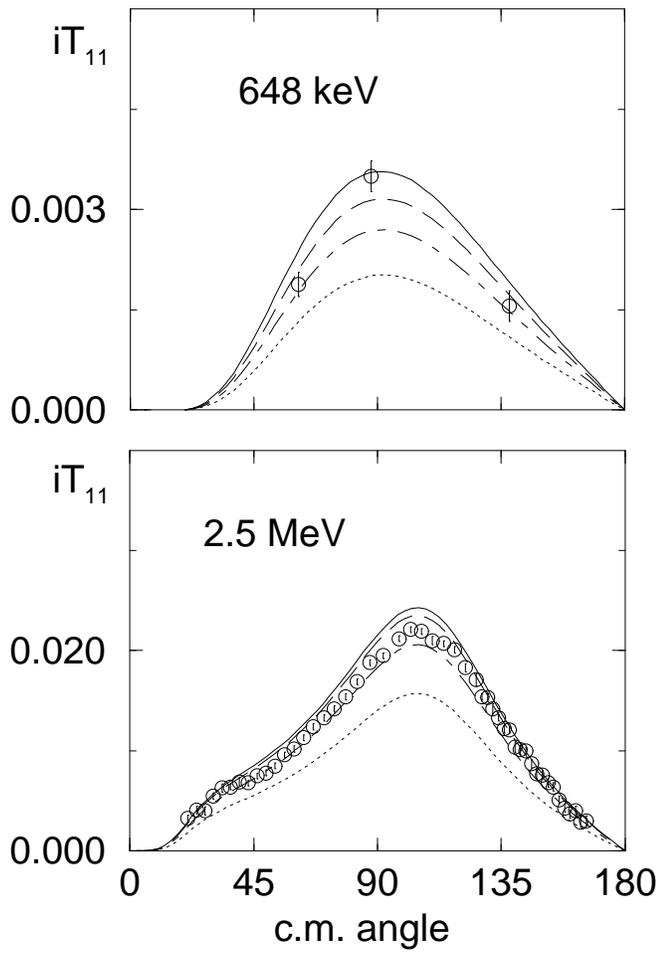
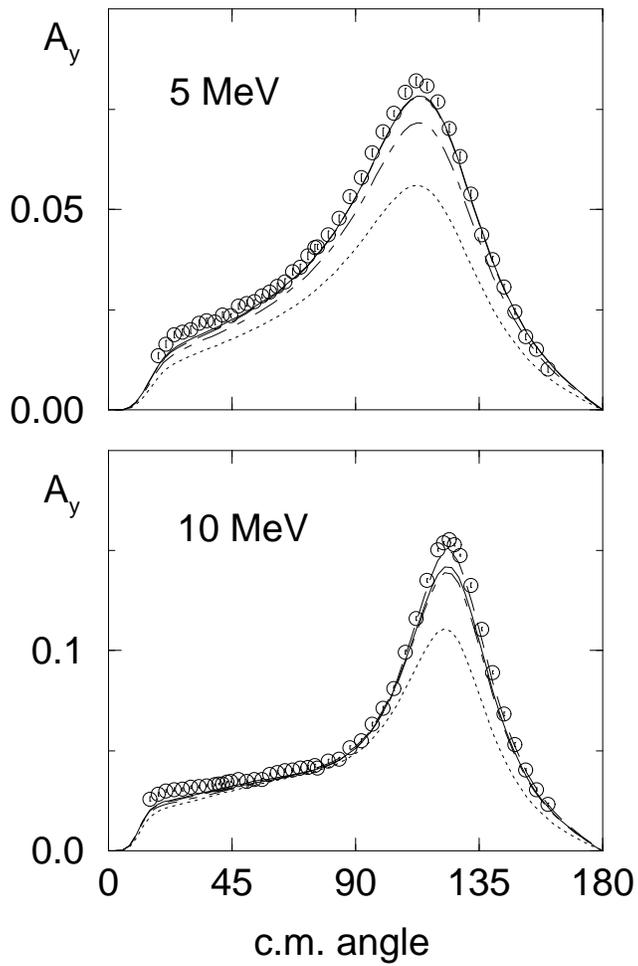
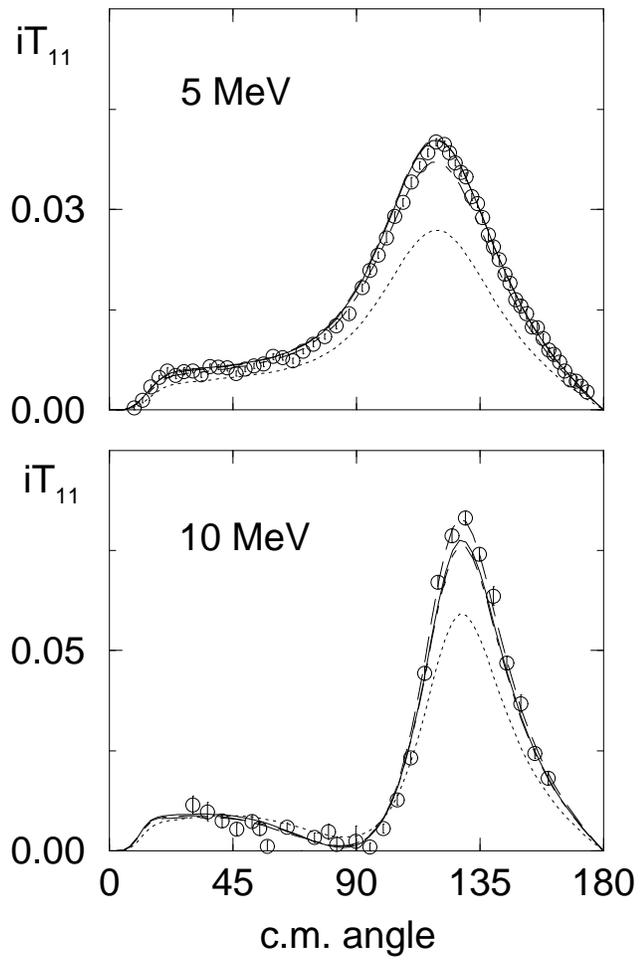
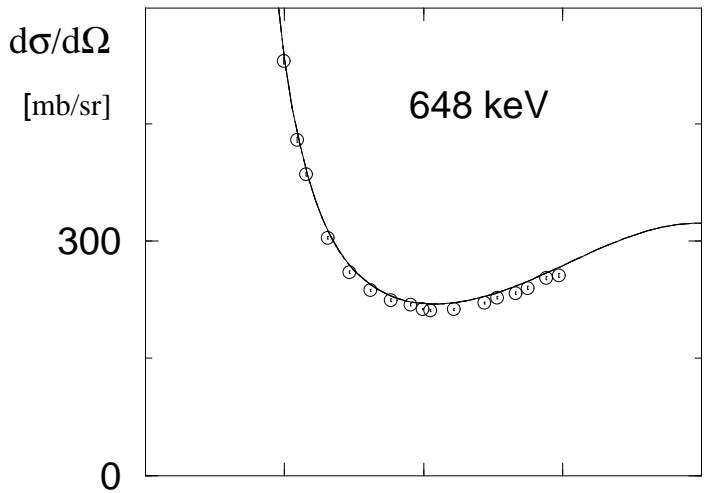
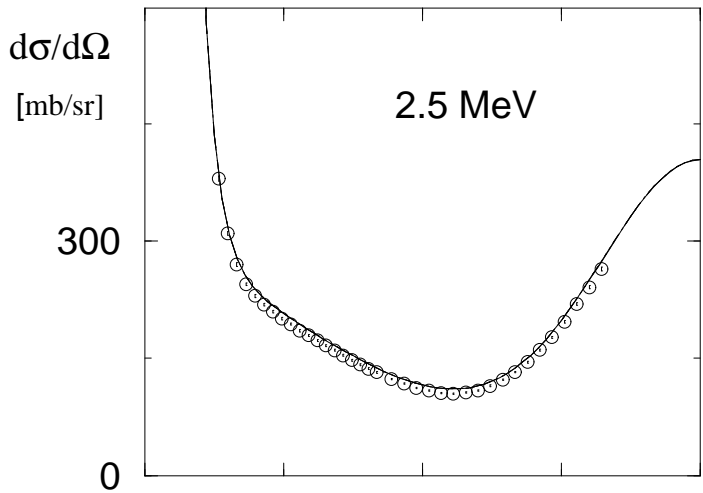
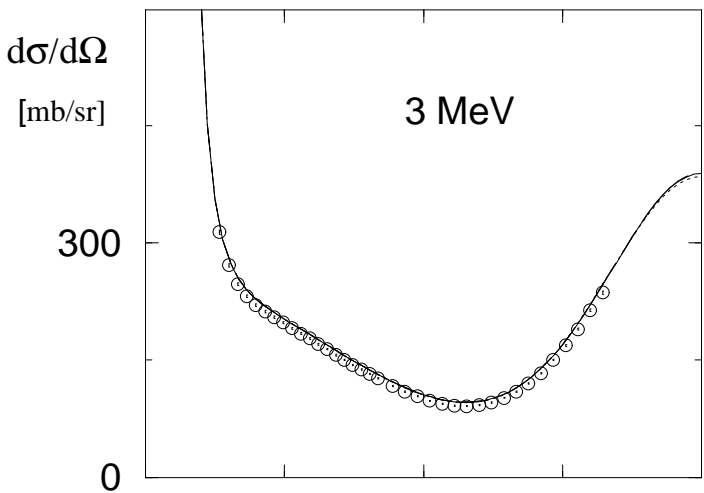
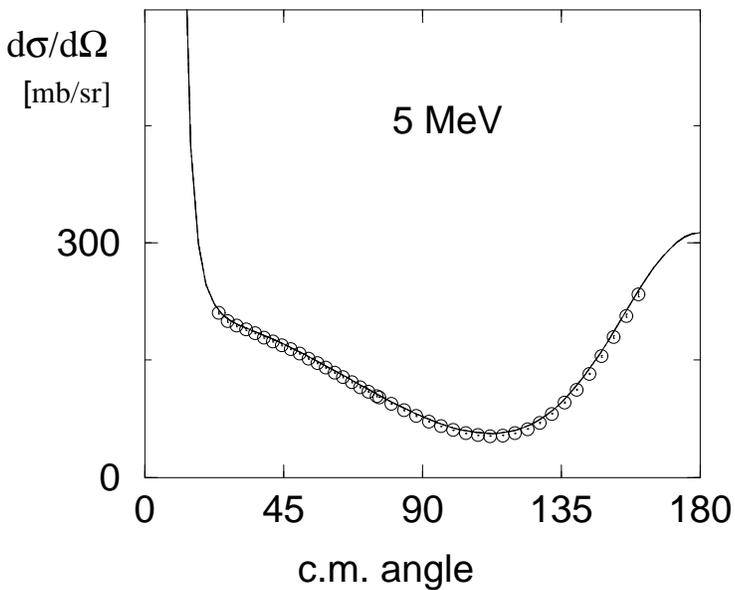
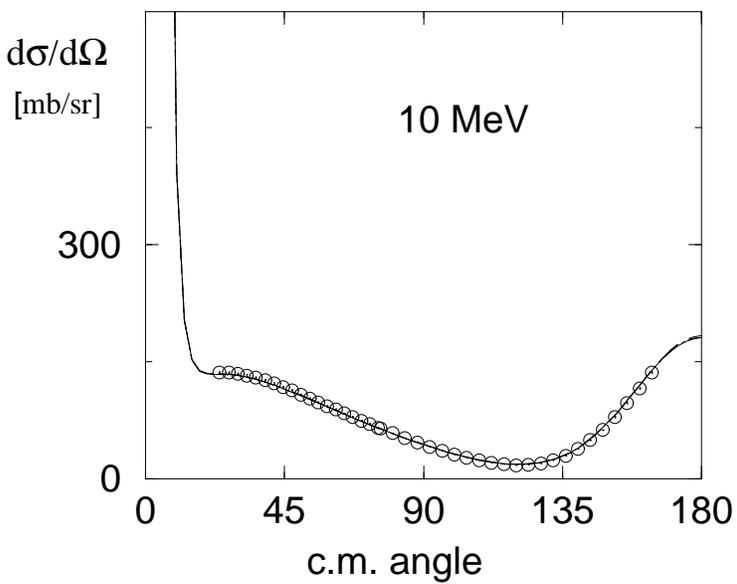
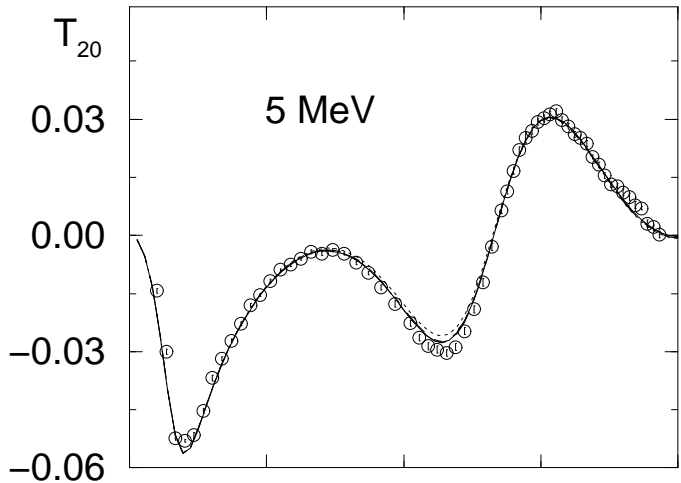
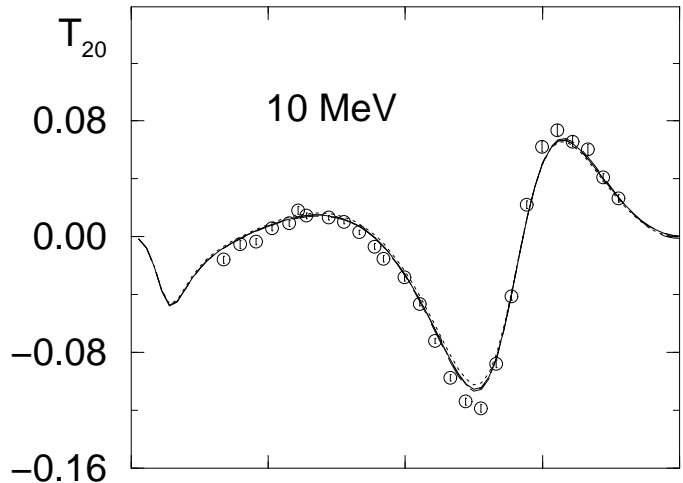
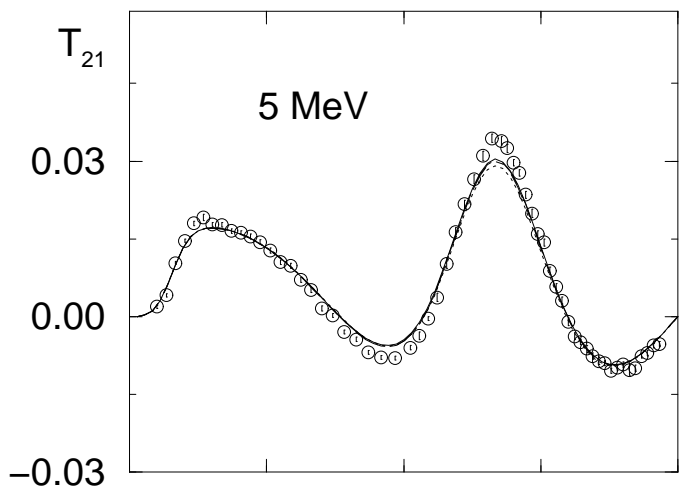
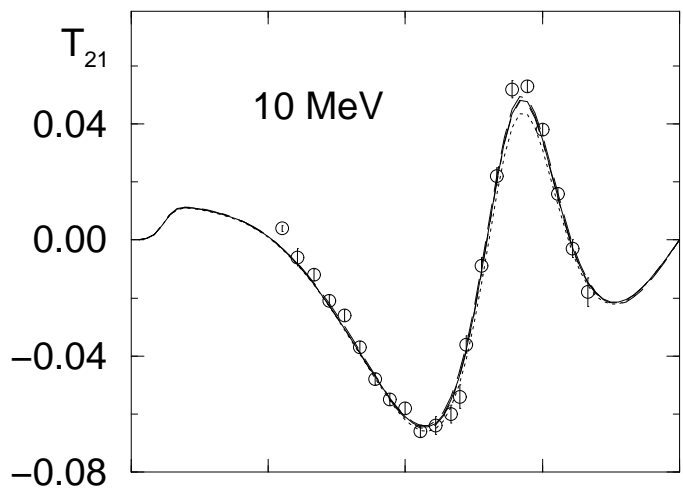
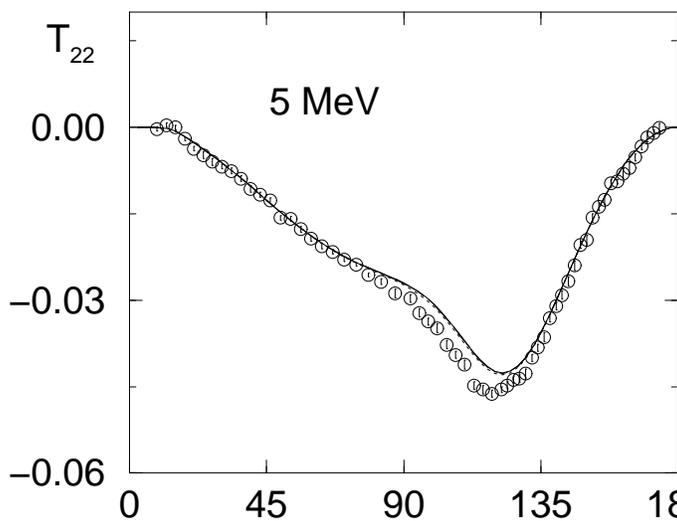
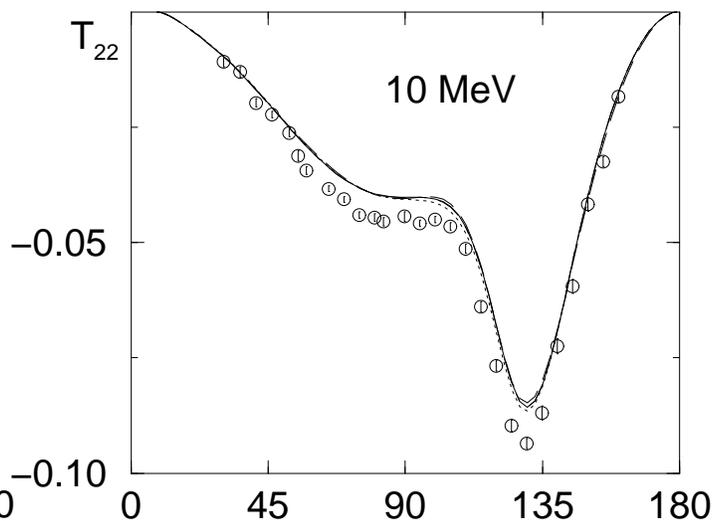
\end{document}